# A holistic multimodal approach to the non-invasive analysis of watercolour paintings

Sotiria Kogou, Andrei Lucian, Sonia Bellesia, Lucia Burgio, Kate Bailey, Charlotte Brooks, Haida Liang

# A holistic multimodal approach to the non-invasive analysis of watercolour paintings

Sotiria Kogou[1], Andrei Lucian[1], Sonia Bellesia[2], Lucia Burgio[2], Kate Bailey[2], Charlotte Brooks [3], Haida Liang[1*]

[1]*School of Science & Technology, Nottingham Trent University, Nottingham NG11 8NS, UK*
[2]*Science Section, Conservation Department, Victoria & Albert Museum, South Kensington, London SW7 2RL, UK*
[3]*Royal Horticultural Society, Lindley Library, 80 Vincent Square, London SW1P 2PE, UK*
*Corresponding author: haida.liang@ntu.ac.uk

## Abstract

A holistic approach using non-invasive multimodal imaging and spectroscopic techniques to study the materials (pigments, drawing materials and paper) and painting techniques of watercolour paintings is presented. The non-invasive imaging and spectroscopic techniques include VIS-NIR reflectance spectroscopy and multispectral imaging, micro-Raman spectroscopy, X-ray fluorescence spectroscopy (XRF) and optical coherence tomography (OCT). The three spectroscopic techniques complement each other in pigment identification. Multispectral imaging (near infrared bands), OCT and micro-Raman complement each other in the visualisation and identification of the drawing material. OCT probes the microstructure and light scattering properties of the substrate while XRF detects the elemental composition that indicates the sizing methods and the filler content. The multiple techniques were applied in a study of forty six 19[th] century Chinese export watercolours from the Victoria & Albert Museum (V&A) and the Royal Horticultural Society (RHS) to examine to what extent the non-invasive analysis techniques employed complement each other and how much useful information about the paintings can be extracted to address art conservation and history questions. A micro-destructive technique of microfade spectrometry was used to assess the vulnerability of the paintings to light exposure. Most of the paint and paper substrates were found to be more stable than ISO Blue Wool 3. The palette was found to be composed of mostly traditional Chinese pigments. While the synthetic pigment, Prussian blue, made in Europe, was found on some of the paintings, none was found on the RHS paintings accurately recorded as being between 1817-1831 even though it is known that Prussian blue was imported to China during this period. The scale insect dyes, lac and cochineal, were detected on nearly every painting including those that fall within the identified date range. Cochineal is known to have been imported to China in the same period. While carbon-based ink was used for the drawings in the V&A paintings, graphite pencil was used in most of the RHS paintings. The majority of RHS paintings were on western papers but nearly all of the V&A paintings were on Chinese papers. Nearly all of the V&A painting substrates and hardly any of the RHS paintings were sized with alum. Elements such as Cu, Zn, Ti and Pb were detected on nearly all of the RHS papers regardless of whether they were Chinese or western papers. These elements were largely absent from the V&A papers. The differences between the two collections reflect their discrete origins and intended purposes.

Keywords: multispectral imaging; reflectance spectroscopy; Raman spectroscopy; X-ray fluorescence spectroscopy; optical coherence tomography; microfade spectrometry

# 1. Introduction

It is well known in a number of disciplines ranging from biomedical science to art conservation and forensic science that no one analytical technique can solve all the problems. This is particularly true in the study of historical paintings where the materials involved are wide-ranging and heterogeneous. A multimodality approach offers new information by combining the strength and overcoming the limitations of individual techniques. For easel and wall paintings, there is a long established tradition of combining a range of different analytical techniques (e.g. optical microscopy, SEM, EDX, FTIR, HPLC etc.) to examine a tiny sample removed from a painting. Paper-based works of art such as watercolour paintings are fragile and the paint layers are rather thin making it impossible to take samples without causing significant damage, which is why most institutions do not allow samples to be taken from paper-based objects. Consequently, invasive scientific analysis can only be conducted on pigment particles which have become detached from the paintings. This has the complication that it is often not possible to know where exactly the pigment particles came from. The advantage of non-invasive analysis is not just in minimising damage, but more importantly in the potential to give a more representative view of the whole object since non-invasive measurements can be taken from anywhere on the object, while samples, if allowed at all, are usually restricted to areas of damage or near the edges.

With the increased availability of mobile instruments for non-invasive examination, it is becoming ever more popular to apply multiple non-invasive techniques for in situ pigment and binder identification in a range of paintings from murals to miniatures. MOLAB has used multimodal non-invasive spectroscopic techniques ranging from X-ray fluorescence spectroscopy (XRF), Raman spectroscopy (532 nm and 785 nm excitations), FTIR (12500 to 900 $cm^{-1}$), UV-VIS reflectance and fluorescence for in situ identification of pigments in easel paintings and binders in wall paintings and contemporary paintings [1]. One of the main challenges for pigment identification in easel paintings as demonstrated in their work was for those pigments containing oxides or sulfides of heavy metals since the FTIR they used was limited at the lower wavenumbers to 900 $cm^{-1}$ due to the use of a chalcogenide glass optical fibre while Raman spectroscopy was usually hindered by the fluorescence of binders and varnishes. Identification of binding media was found to be successful in contemporary easel paintings as they contain a relatively large quantity of binders compared to wall paintings where it was found to be challenging due to the relatively small amount of binders. Miliani et al. [2] compared the effectiveness of material identification using non-invasive spectroscopic techniques (XRF, FTIR, UV-VIS reflectance and fluorescence) with conventional micro-sampling techniques using SEM-EDS, GC-MS, HPLC and FTIR and found the non-invasive techniques performed just as well as the conventional micro-sampling techniques in the identification of inorganic compounds. The non-invasive multimodal techniques were less successful in the identification of organic compounds compared with HPLC and GC-MS on micro-samples. Westlake et al. [3] combined two mobile laser techniques: 1) Laser Induced Breakdown Spectroscopy (LIBS), a micro-destructive technique for in situ and depth-resolved elemental identification and 2) Raman spectroscopy (786 nm excitation) for pigment identification on wall painting fragments. Fluorescence was found to mask the Raman signals in many cases and additionally the 785 nm excitation was also not suitable for copper-containing pigments. Bruni et al. [4] used portable XRF, FTIR, Raman and UV-VIS reflectance spectroscopy to examine paintings and drawings. Aceto et al. [5] suggested an analytical protocol for the identification of pigments in miniatures using non-invasive techniques starting with UV-VIS-NIR reflectance spectroscopy, visual inspection with a high magnification microscope, followed by XRF and then Raman, but found that the main limitation is in the identification of organic colorants and binding media.

One of the major problems with the study of paper-based works of art, such as manuscripts, is the lack of the availability of a wide range of non-invasive instruments to study systematically a large collection of materials [6]. In this project, we address art historical and conservation research questions relating to a large sample of watercolour paintings through the application of a suite of non-invasive imaging and spectroscopic techniques involving multispectral imaging, optical coherence tomography (OCT), X-ray fluorescence (XRF), Raman and VIS-NIR fibre optic reflectance spectroscopy (FORS) to study not only the palette but also the drawing material, painting techniques and the silk or paper substrates (Fig. 1 and Table 1). This is the first time that new techniques such as OCT has been employed in such a large scale study using a multimodal approach.

The Victoria and Albert Museum (V&A) and Royal Horticultural Society (RHS) have large collections of Chinese export watercolour paintings from the $18^{th}$ and $19^{th}$ century [7-8]. The V&A collection comprises a diverse assortment of watercolours on paper and silk which were acquired piecemeal largely from antiquarian booksellers and art dealers in the late $19^{th}$ century. Details of the original collectors and collection dates are seldom available. Subjects include trades and occupations, processions and punishments, shipping, domestic scenes and flower and bird paintings. By contrast, the selected RHS collection of botanical watercolours on Western and Chinese papers was commissioned and collected by an East India tea inspector, John Reeves, between 1817 and 1831. This collection was especially made for the Society and is well-documented. Until this study there has been little scientific analysis conducted on the collections and the descriptions in literature have been largely based on visual inspection. This study aims to better understand the painting techniques and materials of these watercolour paintings, exploring trade and cultural exchanges between China and Europe of this period through insights from scientific analysis [9].

A total of 26 paintings from the RHS and 20 paintings from the V&A were included in this pilot study. They represent a large statistically significant sample which has been studied in a way rarely done previously. The aim of this paper is to examine to what extent the non-invasive analysis techniques employed complement each other and how much useful information about the paintings can be extracted to address conservation and history questions.

## 2. Materials & Methods

An overview of the analysis techniques and the examination strategy is given in Fig. 1. The detailed operational parameters of each technique are summarized in Table 1. In this section, we will describe the instruments, techniques and the sequence of their application.

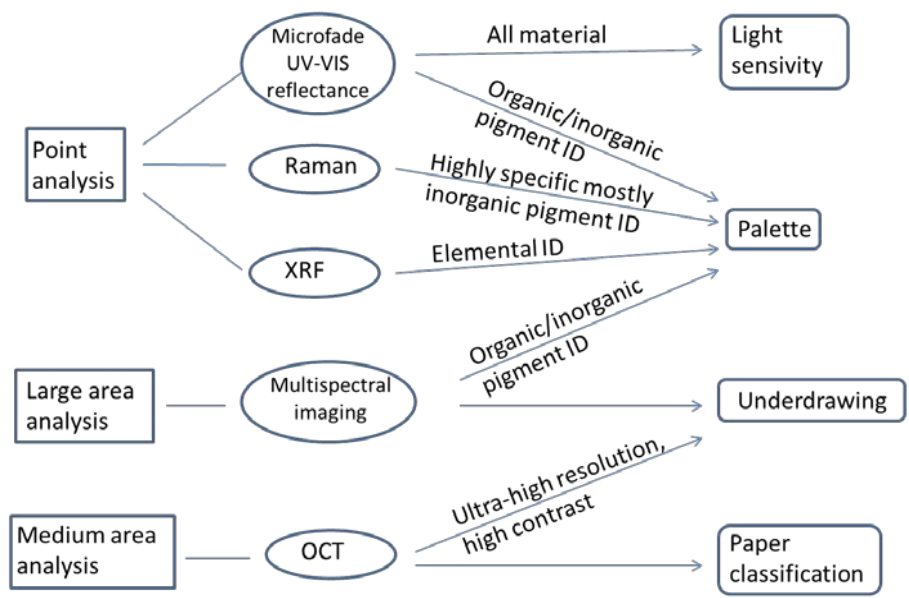

Fig. 1 Overview of the multimodal analysis strategy

## 2.1 *Microfade spectrometry*

The first step was to assess the vulnerability of the material to light exposure and hence establish the appropriate lighting levels for display and analysis since all the proposed non-invasive analysis involved illumination by light. Microfade spectrometry is a micro-destructive technique that allows in situ accelerated light ageing on an object [10]. An in-house built portable fully automatic microfade spectrometer [11] was used for in situ micro-accelerated ageing test of selected spots that represented the range of colours in the paintings. A filtered Xenon light (400-700 nm) was focused to a spot of 0.46 mm × 0.76 mm. The total power over the focussed spot was ~2 mW giving an intensity of ~$2\times10^6$ lux. The light was used both for accelerated ageing and as the illumination for the reflectance spectroscopy which monitored the degradation process. The maximum illumination time was set to 10 minutes but the system automatically cut the light source off as soon as it detected significant change. Since the instrument is more sensitive than the human eye, no damage is noticeable even when the irradiated area is viewed under a microscope.

## 2.2 *Multispectral imaging*

The in-house built versatile remote spectral imaging system, PRISMS (Portable Remote Imaging System for Multispectral Scanning), for efficient high spatial resolution large scale multispectral imaging [12] at distances up to tens of metres, was used for an initial examination of the paintings [13]. Its configuration in the VIS-NIR (400–880 nm) regime consists of a filter wheel with 10 bandpass filters each with a bandwidth of 40 nm except for the one at 880 nm which has a bandwidth of 70 nm, and a thermoelectrically cooled Jenoptics CCD camera. Given that most painting materials have broad spectral features in the VIS-NIR range, the spectral resolution of the PRISMS system was sufficient for the identification of most pigments of interest. However, the low spectral resolution precludes the accurate recording of the fine spectral features (fingerprints) in the anthraquinone dyes such as madder (plant dyes), cochineal and lac (scale insect dyes) (Fig. 12) and cobalt pigments such as smalt (Fig. 10). PRISMS was therefore used for preliminary examination of the paintings to reveal the underdrawing using the near infrared bands, grouping areas with similar spectra and defining the different areas of interest to be analysed using the point analysis methods.

2.3 *High spectral resolution VIS-NIR reflectance spectroscopy*

For detailed pigment identification where a higher spectral resolution and/or spectral sampling than that offered by PRISMS is needed. FORS in the form of the microfade spectrometer with a low intensity tungsten light source (Ocean Optics DH-2000) was used. The spectral range was 400-950 nm, the spectral resolution and spectral sampling were 0.9 nm and 0.5 nm, respectively. For the anthraquinone dyes and cobalt pigments where the highest spectral resolution and spectral sampling are required, the optimum signal-to-noise spectra that contained all the spectral features without distortion were obtained with a smoothing window of width 10 nm.

Two Polychromix fibre optic spectrometers were used for the short wave infrared (SWIR) spectral range (1000-2500 nm): DTS 1700 (900–1700 nm) and DTS 2500 (1700–2500 nm) with spectral resolutions 12 and 22 nm respectively. Additional indicative spectral absorption features can be found in this spectral range for pigments such as lead white, azurite, malachite, atacamite and cobalt pigments. The spectral range is also known to give information on the binding media allowing the separation of triglyceride-containing binders such as egg yolk and oil from gum arabic [14] and sometimes animal glue [15]. Preliminary examination of paper showed that the strong spectral absorption features of paper in the SWIR range made the identification of the binders challenging. Further investigation showed that the absorption features due to the paper substrate masked the characteristic spectral features of most pigments in this spectral range. Prussian blue is perhaps the only pigment that can be identified using the SWIR spectral range, as it has a characteristic positive slope between 1000 and 1300 nm.

It is commonly believed that while FORS measurements can identify single pigments, it is less effective for the identification of mixtures of pigments [16]. In the following sections, we will demonstrate that it can be as effective in identifying mixtures of pigments using an algorithm that is based on the Kubelka-Munk model.

2.4 Pigment identification using the Kubelka-Munk model

Given that most of the colorants in these paintings are either in layers or mixtures, a Kubelka-Munk (KM) based method was used for pigment mixture identification using the spectral reflectance measurements [12]. The simplest form of KM model is given by $\frac{K}{S} = \frac{(1-R_\infty)^2}{2R_\infty}$, where $R_\infty$ is the spectral reflectance of a layer with infinite optical thickness and $K$ and $S$ are the spectral dependent effective absorption and scattering coefficients respectively. KM theory works best for highly scattering paint where the scattering particle size is smaller than the thickness of the paint. In other words, it does not work so well for transparent paint or paint that is high in absorption ($K$) but low in scattering ($S$) coefficient (e.g. Fig. 3). The assumption that K and S scales linearly with concentration is often a good approximation as long as the concentration is not too high. If we assume that the absorption and scattering coefficients of a paint mixture is a linear combination of the K and S of the individual pigment components, then we can use KM theory to predict the reflectance of the paint mixture. For pigment identification, we can set the concentration of the single pigments as free parameters and find the best non-negative least squared fit of the predicted reflectance spectrum to the measured spectrum of the mixture. A number of indicators for goodness of fit can be defined. For example, the cross-correlation coefficient calculated from the predicted and actual spectra of the mixture is a good indicator on how closely the peaks and absorption troughs match in position. This indicator is well suited to pigment identification since in this case it is the match in spectral characteristics and not so much the absolute match of the spectra that is important. Once the maximum number of components is defined, the algorithm then finds automatically the best combination of pigments from the reference spectral library that fits a measured spectrum. Alternatively, the KM model can be used to test hypotheses. For example, it can be used to check if the pigments identified by Raman and XRF matches the measured VIS-NIR reflectance spectrum.

The effect of pigment concentration on KM fit was examined. Solutions of each pigment at 25% mass concentration were prepared and the concentration variation was controlled through the number of layers applied (1, 3 and 10 layers for low, medium and high concentration respectively). If $K/S$ scales linearly with concentration, then a change in concentration would result in a shift up or down by a constant in the $-\log(K/S)$ versus wavelength plot without changing the spectral shape. For most of the reference pigments, $-\log(K/S)$ derived using the KM model from the reflectance measurements, seem to have a more or less linear dependence on concentration (Fig 2a). For these pigments, the spectrum of low pigment concentration can produce good fits to a high concentration paint-out. On the other hand, there are pigments like gamboge that show spectral shifts as large as ~50nm and spectral feature changes from extremely low to extremely high concentrations (Fig. 2b). In these cases, the use of low concentration reference cannot provide a good fit to areas where high concentration pigment has been used. However, even in these cases, reasonably good fits for a wide range of intermediate concentration samples can be obtained as long as both a high and a low concentration reference are available in the reference spectral library.

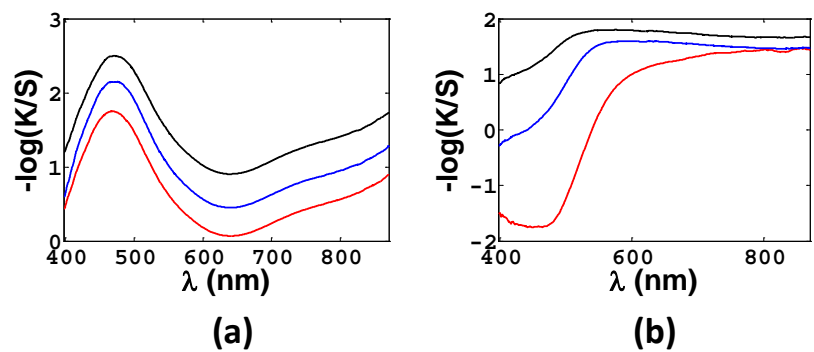

Fig 2. The effect of pigment concentration on $-\log(K/S)$ plot of a) azurite (grade 5) and b) gamboge in animal glue at 25% mass concentration applied on watercolour paper with different number of layers (red curve: 10 layers, blue curve: 3 layers and black curve: single layer).

Pigment particle size can affect the colour appearance and this property is used in Chinese paintings to produce different colour saturations without the addition of white pigments. Coarse grains of azurite and malachite produce more saturated colours than finer grains. Pigment particle sizes can also affect the spectral shape resulting in shifts in spectral features [12,17]. The shifts and alterations in spectral features due to differences in concentration and particle size have to be taken into account in pigment identification and the reference samples should cover all these cases.

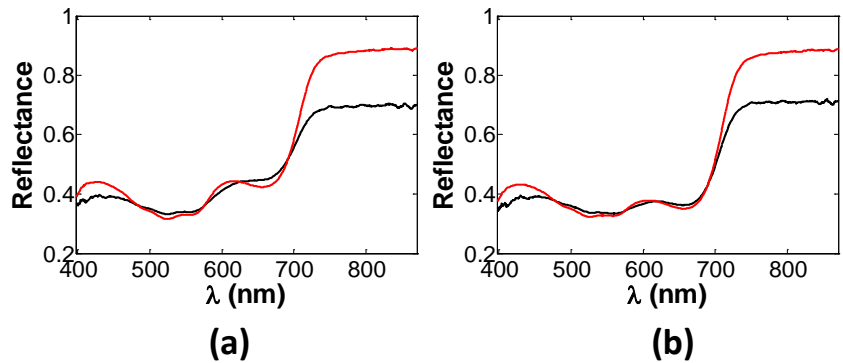

Fig 3. Reflectance spectra (black curve) of a) a mixture and b) a layer-by-layer application of cochineal lake and indigo fitted with the KM model (red curve) using the reference spectra of the two single pigments.

It is a common technique in Chinese paintings to use different colour paint in layers in order to achieve the desired effect [17-18]. The KM model is known to be designed for mixtures of pigments, however, it was found to work well even for pigments painted in layers in the case of watercolours. The paint layers are relatively thin in watercolour and therefore the difference between mixed and layered paints is relatively small. A series of experiments were conducted using different types of pigments painted in layers, and the KM method was shown to provide equally good fits in spectral features for areas where pigments have been applied in layers (Fig. 3). The poorer fit at wavelength >700nm is because indigo absorbs strongly below 700nm but becomes highly transparent above this wavelength, while cochineal is highly transparent over most of the spectral range. KM model do not work so well for low scattering material. However, this does not affect pigment identification as it is the spectral features rather than the relative amplitudes that is important.

The effectiveness of the KM method on samples that combine both mixture and layer-by-layer application was also examined. They showed that good fits can be achieved in these cases as well. Figure 4 shows KM fits of the spectra of green areas in which a mixture of gamboge and indigo was applied on top of a malachite layer. While the fact that KM model is insensitive to whether the paints are in mixtures or layers is convenient for pigment identification, it also means that the method cannot be used to distinguish between mixtures and layers for watercolours. However, careful examination of high resolution images can give clues as to whether it is painted in layers or in mixtures. In some cases, where the paint is applied thick, it is possible to see the layers using OCT (Section 2.7).

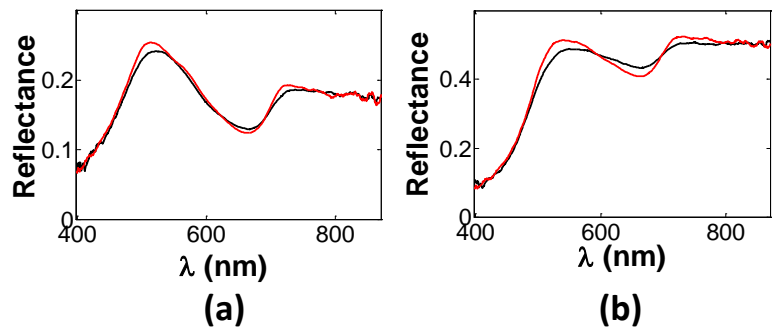

Fig 4. Reflectance spectra (black curve) of a a) dark and b) a light green area painted by applying a mixture of gamboge and indigo on top of a layer of malachite fitted with the KM model (red curve) using the reference spectra of the 3 single pigments.

2.5 *X-ray fluorescence spectroscopy (XRF)*

An ArtTAX XRF spectrometer was used to identify the elemental content of the painting materials and substrates. The operational parameters are 50 kV, 600 μA and 100 seconds of accumulation live time. Since it was used in open air without helium purge, it was normally only sensitive to elements with atomic number $Z > 14$ which was an obvious limitation. However, complementary techniques that provide molecular information such as VIS-NIR reflectance spectroscopy and Raman spectroscopy when used in combination can overcome or at least partially compensate for this limitation.

2.6 *Raman spectroscopy*

Raman spectroscopy provides highly specific molecular identification and can even distinguish between different crystalline phases of compounds with the same molecular formula [19]. The increased availability of mobile and handheld Raman devices have also encouraged their application in heritage science [20]. In this project, micro-Raman analysis was performed with a Horiba XploRA spectrometer coupled to an Olympus microscope using a 638 nm and a 532 nm laser. The laser beam was focussed to a ~2 μm spot allowing the analysis of individual particles and limiting the interference from surrounding materials. Micro-Raman spectroscopy is well-suited to the identification of inorganic pigments but its usefulness in the identification of organic materials is usually limited by fluorescence [21]. It can nevertheless detect some organic materials such as indigo and gamboge. Its use on easel paintings is restricted by the high fluorescence of varnishes and binders. However, its use on paper based works of art where water-based binding media have been employed is in general more successful. The inability of Raman microscopy at detecting most organic dyes was partially overcome by using VIS-NIR reflectance spectroscopy which is highly sensitive to thin washes of dyes. Raman is also a surface analysis technique, usually restricting the identification of materials to the top layer of the area examined.

2.7 *Optical Coherence Tomography*

Imaging in the NIR is usually used to detect preparatory sketches under paint layers since most paints are relatively transparent in the NIR whilst carbon based drawings absorb strongly in the same spectral range. The NIR bands of the multispectral imager PRISMS can be used for large scale survey of preparatory sketches. However, a relatively new technique, Optical Coherence Tomography (OCT) is known to provide superior images of underdrawings [22-23]. OCT is based on an imaging Michelson interferometer which provides rapid non-contact and non-invasive cross-section images of subsurface microstructures. It provides the highest contrast and resolution images of underdrawings than any other imaging modality owing to its high resolution and the possibility of choosing the slices in the depth range that contained the best images of the underdrawings.

OCT has been used to characterise industrial paper, measuring the microstructure and the filler content of paper using the slope of the depth profile of the backscattered light intensity [24]. In this project, OCT was used to distinguish the different types of papers both in terms of the length of the fibres and the scattering properties of the papers.

The OCT used is an adapted Thorlabs spectral domain OCT at 930 nm with axial resolution of ~7 μm in air (or ~4.5 μm in paint, cellulose and silk) and transverse resolution of 9 μm.

Table 1 Spatial and spectral parameters of each technique

| Technique | Working distance (mm) | Spatial resolution or spot size (mm) | Field of view or scanned size (mm) | Spectral range |
|---|---|---|---|---|
| Micro-Raman | ~0.5 | ~0.002 | ~0.002 | 532 nm, 638 nm excitation, Raman shift range 150-3800 $cm^{-1}$ |
| XRF | ~1 | ~0.2 | ~0.2 | 2-50 keV |
| Microfade | ~30 | ~0.46×0.76 | ~0.46×0.76 | 400-700 nm |
| VIS-NIR FORS | ~30 | ~0.5 | ~0.5 | 400-950 nm |
| SWIR FORS | ~2 | ~5 | ~5 | 900-2400 nm |
| UV-VIS-NIR multispectral imaging | 2240 | ~0.085 | ~115×85 | 400-900 nm |
| OCT | ~10 | ~0.009×0.009×0.0045 | ~10 (width) | 930 nm (100 nm bandwidth) |

2.8 Reference Samples

Various considerations were taken into account when preparing the reference mock up samples for pigment identification. Firstly, a variety of different pigments that were traditionally used in China and some pigments and dyes used in Europe in the period of interest were selected. The lake pigments were prepared from the dyestuff by J. Kirby at the National Gallery (London) laboratory [25-27]. Table 2 lists the reference pigments and their composition, most of which were analysed and verified in the laboratory in a previous project [25]. The binder was animal glue from the Pigment Factory in Beijing China. The paints were applied on two substrates: Chinese Xuan paper and Canson Aquarelle-300gm watercolour paper. The preliminary study in section 2.4 suggested the need to include both high and low concentration paint-outs for each pigment and a range of different particle sizes for those mineral pigments that are traditionally graded into coarse, medium and fine particle sizes. A VIS-NIR spectral library was compiled from the reference samples using FORS as described in section 2.3. Standard spectral reference libraries [28-29] were used for Raman analysis.

Table 2 Reference pigments and composition

| Pigment name and supplier | Composition (major components) |
|---|---|
| ***Red*** | |
| Iron oxide red, natural: Kremer Pigmente | Natural red earth rich in iron oxide |
| Red lead: Kremer Pigmente | $Pb_3O_4$ |
| Vermilion light: Kremer Pigmente | HgS |
| Lac | Lac dye: sticklac (Kremer Pigmente) soaked in warm water and filtered using cotton patches. |
| | Lac Al-based lake (NG laboratory) |
| Cochineal | Cochineal dye: cochineal insects (Kremer Pigmente) soaked in warm water and filtered using cotton patches. |
| | Cochineal Al-based lake (NG laboratory): 18$^{th}$ century type |
| | Cochineal Sn-based lake (NG laboratory): 19$^{th}$ century type with $SnCl_4$ |
| Madder lake: NG laboratory | Madder dyestuff on Al substrate: 19$^{th}$ century type |
| Rose madder: L. Cornelissen & Son | Madder dyestuff on a sulphate-containing alumina substrate, 19$^{th}$ century type |
| Sappanwood lake: NG laboratory | |
| Crimson alizarin: Roberson & Co | Alizarin on a sulphate-containing alumina substrate |
| Rouge: Pigment Factory Beijing | Organic dyes, composition unknown |
| | |
| ***Yellow*** | |
| Realgar (Grade 3): Pigment Factory Beijing | $As_4S_4$ |
| Orpiment : Kremer Pigmente | $As_2S_3$ |
| Yellow ochre: Kremer Pigmente | Natural yellow earth |
| Litharge: L. Cornelissen & Son | Massicot (orthorhombic PbO) |
| Gamboge: Pigment Factory Beijing | ready-prepared paints in animal glue |
| Buckthorn lake: NG laboratory | Buckthorn dyestuff on alumina substrate |
| Weld lake: NG laboratory | Weld dyestuff on an alumina substrate |
| Quercitron lake: NG laboratory | Quercitron dyestuff on a Tingry alumina substrate, 19$^{th}$ century recipe |
| | |
| ***Blue*** | |
| Azurite (Grade 1,3,5): Pigment Factory Beijing | $2CuCO_3 \cdot Cu(OH)_2$ |
| Indigo: Pigment Factory Beijing | $C_{16}H_{10}N_2O_2$ ready-prepared paints in animal glue |
| Indigo: Kremer Pigmente | Indigo and silicaceous extender |
| Prussian blue (Milori Blue): Kremer Pigmente | Hydrated iron hexacyanoferrate complex, $KFe[Fe(CN)_6] \cdot H_2O$ |
| Cobalt blue medium: Kremer Pigmente | Cobalt aluminium oxide |
| Smalt, L.Cornelissen & Son | Cobalt-containing potash glass |
| Ultramarine synthetic (dark): Kremer Pigmente | Sulphur-containing sodium aluminosilicate, approx. $Na_{6-10}Al_6Si_6O_{24}S_{2-4}$ |
| | |
| ***Green*** | |
| Malachite (Grade 1,3,5): Pigment Factory Beijing | $CuCO_3 \cdot Cu(OH)_2$ |
| Atacamite: Kremer Pigmente | $Cu_2Cl(OH)_3$ |
| | |
| ***White*** | |
| Lead white (Cremnitz white): Kremer Pigmente | $PbCO_3 \cdot Pb(OH)_2$ |
| Shell white: The Pigment Factory Beijing | $CaCO_3$ |
| | |
| ***Black*** | |
| Chinese ink | Carbon |

## 3. Results

3.1 *Light sensitivity*

All the paints and paper substrates tested with the microfade spectrometer were found to be more stable than ISO Blue Wool 3, except for a yellow paint in one of the V&A paintings which degraded faster than ISO Blue Wool 1. The spectrum, light induced spectral changes and the rate of the degradation of this yellow paint suggested that it was most likely to be realgar. Subsequent analysis with micro-Raman spectroscopy confirmed the identification (Fig. 9). On the whole, the paints are more stable to light than the paper substrate.

3.2 *Multimodal technique for pigment identification*

The complementarity of the three different spectroscopic techniques, VIS-NIR reflectance spectroscopy, XRF and micro-Raman will be examined in the following case studies.

3.2.1 *Single pigments*

It was found that in cases where a single pigment dominates, the results from all three techniques tended to agree. An example of a blue region in a V&A painting is shown in Fig. 5 where XRF detected Cu as the main element, Raman identified individual particles of azurite, malachite and goethite. Visual microscopic examination of the region showed that the blue azurite particles were mixed with the occasional green malachite and brown goethite particles which are the usual impurities of the azurite mineral [30]. VIS-NIR reflectance spectroscopy showed that the spectrum was consistent with azurite (coarse particle grade). As the VIS-NIR reflectance spectroscopy collects light over a larger spot or area than Raman and XRF, it provides statistically more representative information of the blue region.

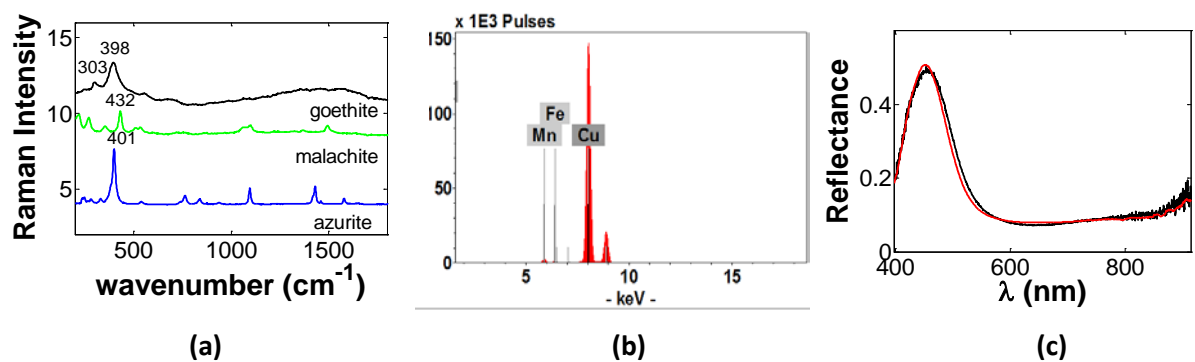

(a) (b) (c)

Fig 5 a) Raman spectra of particles in a blue area on painting no.7791-12 (V&A collection); b) XRF spectrum detected Cu as the main element and c) the best fit to the VIS-NIR reflectance measurement (black curve) using the KM model (red curve) was with a reference spectrum of grade 1 azurite.

*3.2.2 Multiple pigments*

In cases where more than one pigment is involved, identification becomes more complicated and the combined use of these three techniques is required in order to obtain reliable results.

VIS-NIR reflectance spectroscopy cannot give reliable identifications of yellow pigments (except perhaps yellow ochre) as their reflectance spectra in VIS-NIR all have an S shaped spectrum with less than 50 nm difference in the position of the points of inflexion. This small spectral difference is no more than the range of spectral differences in gamboge due to concentration changes (see section 2.4). XRF and Raman are good techniques for the detection of yellow inorganic pigments. For example, chrome yellow was detected in one painting by Raman while XRF found Cr confirming that it was not just signal from an isolated particle since the XRF spot size was much larger (see Table 1).

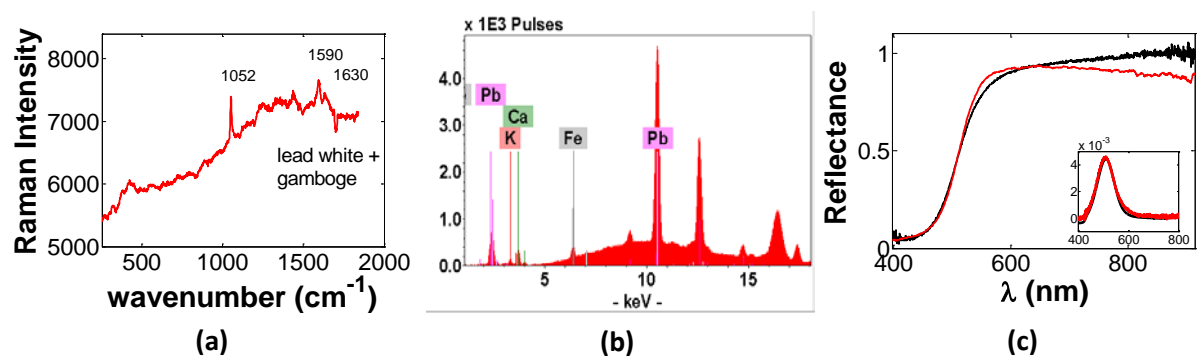

Fig 6 a) Raman spectrum of a particle in a yellow area of painting no.7790-20 (V&A collection) show both lead white and gamboge; b) XRF spectrum detected Pb as the main element and c) KM fit (red curve) to VIS-NIR reflectance spectrum (black curve) consistent with a lead white and gamboge mix. The inset shows the corresponding derivatives of the spectra.

The identification of organic yellow pigments is particularly challenging as XRF is not generally useful for organics and Raman is often hindered by the fluorescence induced by the laser in organic materials. Figure 6 shows the results of the examination of a yellow area where Raman detected lead white and gamboge, XRF detected Pb as the main element and VIS-NIR reflectance spectrum was found to be consistent with a mixture of lead white and gamboge using the KM model. It is significant that Raman detected gamboge with certainty. Gamboge is known to be one of the most commonly used yellow colorants in Chinese paintings, however, its identification on historical East Asian paintings has proved challenging so far [17].

Green leaves are commonly found in these paintings and therefore provide a significant statistical sample for examining the complementarity of our analytical techniques. It also allows the examination of any variations, if any, of the techniques for painting green leaves. In nearly all the areas with green leaves, FORS spectra were consistent with a mixture of malachite, indigo and gamboge with the occasional addition of lead white. The existence of malachite was always supported by the detection of Cu using XRF. XRF is not sensitive to organic pigments like indigo and gamboge. Raman rarely detects all three pigments when they are present together. This is due to a combination of factors: malachite is a weak Raman scatterer and in the case of green leaves it was often painted below a layer of indigo and gamboge and therefore not excited by the laser beam, the two organic dyes can be quite fluorescent, and the small spot size of micro-Raman makes it difficult to obtain a statistically complete view of the pigment composition. In the example shown in Fig. 7, Raman detected lead white and gamboge, XRF detected Cu and Pb as the main elements while the FORS spectrum is best fitted with a combination of gamboge, lead white, malachite and indigo.

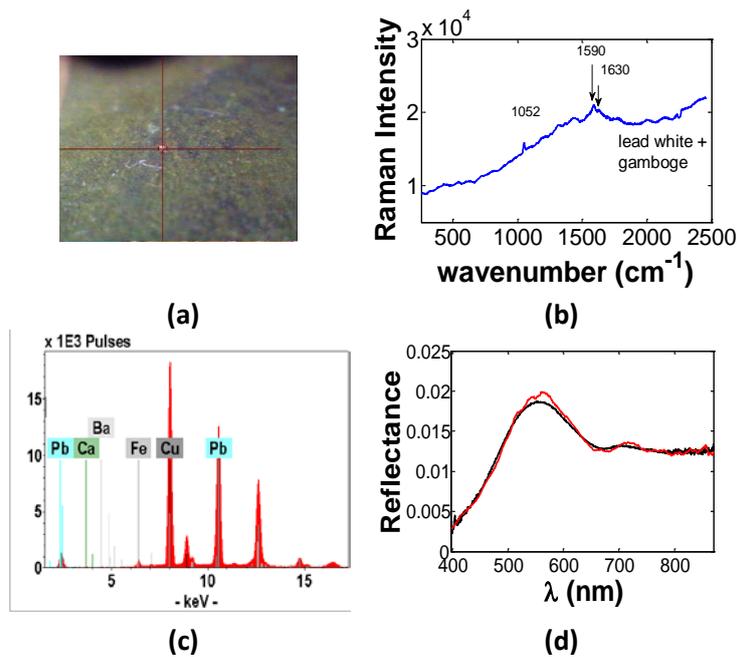

Fig 7. a) Microscopic image of a green area on painting D.291-1886 (V&A collection); b) Raman measurement for this area detected lead white and gamboge; c) XRF spectroscopy found mainly Cu and Pb and d) KM model fit (red curve) using a combination of lead white, gamboge, malachite and indigo to the measured VIS-NIR reflectance spectrum (black curve).

Another area of interest was the detection of smalt on one of the paintings from V&A. The blue region shown in Fig. 8 was found to have in its FORS spectrum the characteristic (narrow) spectral features of smalt while the best fit of the spectrum was with a KM model mixture of smalt, azurite, lead white and Chinese ink. XRF detected Cu as the main element and small amounts of Co and Pb, thus supporting the presence of azurite, smalt and lead white. Raman detected only azurite with impurities of goethite and malachite.

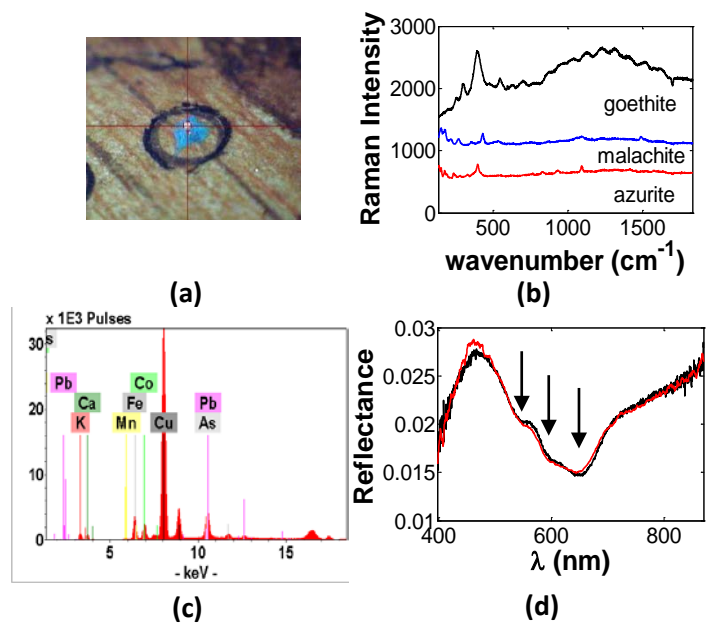

Fig 8. a) Microscopic image of a blue area on painting no.1312-1889 (V&A collection); b) Raman spectra for particles in the blue area; c) XRF spectroscopy detected Cu (main element) with small amounts of Co and Pb. d) KM model fit (red curve) to the measured FORS spectrum (black curve) is consistent with a mixture of smalt, azurite (grade 5), lead white and Chinese ink. The arrows indicate the positions of the characteristic absorption lines of smalt.

The examination of degraded yellow areas containing arsenic was interesting. The pigment was first suspected to be realgar because of its degradation characteristics under microfade (Section 3.1). Figure 9 shows that XRF found As and Pb to be the main elements while Raman analysis of particles from this area showed that it is a mixture of two phases, both of which are intermediate between realgar and pararealgar (both a polymorph and a degradation product of realgar). Realgar and pararealgar along with those intermediate phases co-exist in the natural mineral and therefore these phases may have been present even in the original painting material. Raman is the only one of the three techniques capable of discriminating between all these phases. As expected, the FORS spectrum is not very well fitted with a standard realgar spectrum with or without the addition of lead white.

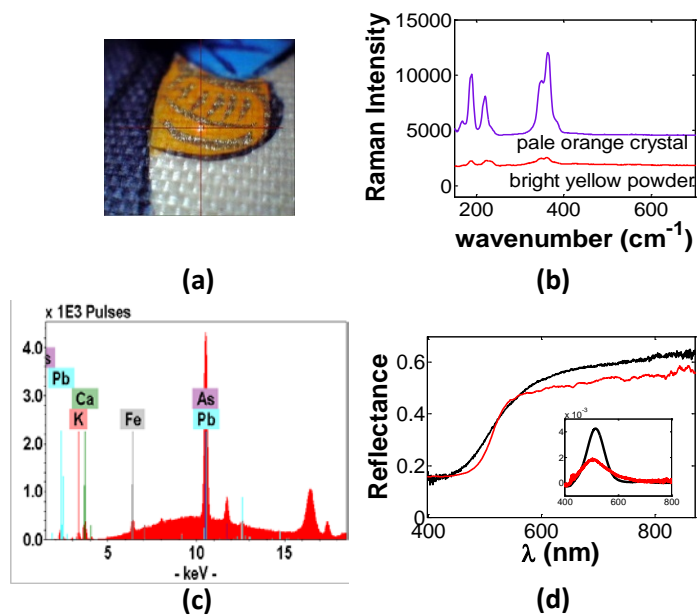

Fig 9. a) Microscopic image of a yellow area on painting no. 7790-8 (V&A collection); b) Raman measurement identified a mixture of two phases both of which are intermediate between realgar and pararealgar; c) XRF spectroscopy detected Pb and As as the main elements; d) KM fit with a mixture of realgar and lead white (red curve) to FORS measurement (black curve); The inset shows the corresponding derivatives of the spectra.

The importance of using these complementary techniques is further illustrated in areas containing scale insect dyes. These organic dyes fluoresce strongly under Raman analysis. In the example illustrated in Fig. 10, Raman detected vermilion along with a high fluorescence signal suggesting the presence of organic materials. XRF detected Pb as the main element along with small amounts of Hg and Fe. The Fe may be from the paper substrate as it also contains Fe. FORS detected the characteristic narrow absorption features of scale insect dyes between 500 and 600 nm.

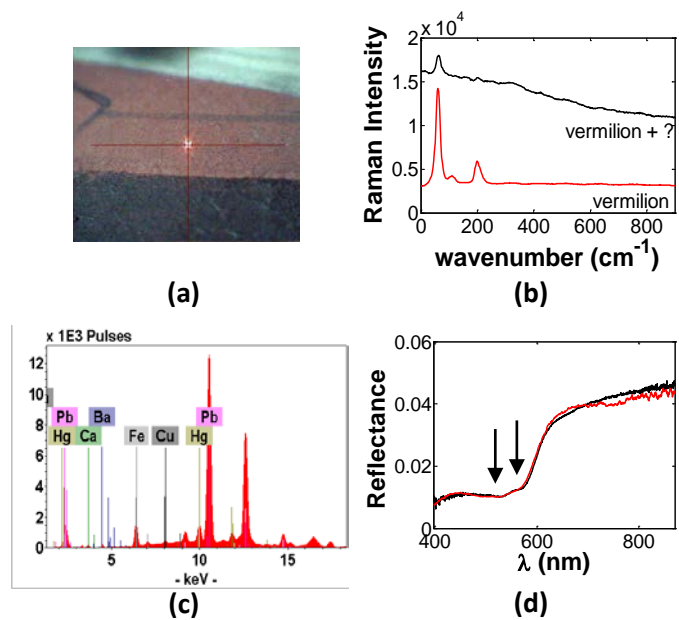

Fig 10. a) Microscopic image of a red area on painting no. D82-1886 (V&A collection); b) Raman spectrum (black curve) compared with a standard Raman spectrum of vermilion (red curve) showed that the examined area is likely to consist of vermilion and a fluorescing material; c) XRF detected mainly Pb and Hg; d) KM fit (red curve) to FORS measurement (black curve) using a mixture of lead white, vermilion and a lac dye. The arrows indicate the positions of the 2 characteristic absorption lines of lac dye.

The identification of the scale insect dyes deserves a more in-depth investigation. Scale insect dyes such as lac and cochineal belongs to the class of anthraquinone dyes which also includes madder. Lac, cochineal and madder are of particular interest in the context of this project because it is known that lac and madder had been used in China over the centuries but cochineal was imported to China during the period of interest [17-18]. The definitive non-invasive identification of these colorants is particularly challenging, especially when they are applied in a thin wash, as they are highly fluorescent when analysed by Raman and they produce very weak signal in FTIR spectroscopy [31]. VIS reflectance spectroscopy is one of the most sensitive non-invasive techniques for the detection of anthraquinones dyes [32-33]. The two narrow absorption lines between 500 and 600 nm for madder are generally at shorter wavelength than those for lac and cochineal, and they can be used to distinguish madder from the scale insect dyes [34-35]. It has also been noted that these dyes are sensitive to the pH of their environment. In experiments carried out in buffer solutions of a range of pH values, it was found that the characteristic absorption lines of cochineal and lac were less prominent and were blue shifted in acidic solutions [36-37]. Colour changes were also observed when cochineal was painted on paper of different pH values [38]. UV-fluorescence spectroscopy is also one of the common techniques for detecting organic dyes, but the broad emission features in roughly the same spectral range for the red dyes along with the complication of the self-absorption effects [27] and fluorescence of the binding media and substrates makes the technique less effective [31].

Scale insect dyes were detected in nearly all the paintings from the two collections using FORS based on the position of the two absorption features between 500 and 600 nm. It is not clear from the literature on Chinese paintings whether it was the dyes alone and/or the lake pigments made from the dyes that were used. In any case, for the period of interest, we cannot assume that European lake pigments were not exported to China. In Europe, scale insect dyes have been used in both forms, though mainly in the form of lake pigments in easel paintings. Lake pigments made from these dyes (Table 2) were painted out in animal glue on untreated Xuan paper. Al and Sn-based cochineal lakes and alum-based lac lakes prepared by the National Gallery in London were examined. The Al-based cochineal lake was prepared following an 18$^{th}$ century recipe using alum, while the Sn-based cochineal lake was prepared following an early 19$^{th}$ century recipe [26]. Given that the position of the absorption lines of cochineal and lac dyes can shift according to the pH value, it was decided that realistic mock-ups were needed to mimic the likely range of pH values in such paintings. Reference samples using these dyes were prepared on Chinese Xuan paper in the traditional manner, where the paper substrate was first brushed with a solution of alum and animal glue and then painted with layers of dyes in animal glue followed by solutions of alum and animal glue and the process was repeated 3 times to achieve a stronger colour [17-18]. The solution of alum and animal glue was used to fix the paint [18]. A subtle wavelength shift of a few nm between the absorption features in the lac and cochineal dye samples was found. The accuracy in the determination of the absorption line positions is ~1 nm. The absorption lines of the madder sample are about 10 nm shorter than those of lac and cochineal and therefore easily identified. Madder was not detected by FORS in any of the paintings.

XRF measurements of the reference samples showed that S was detected in the Al-based lake pigments as expected since alum which contains sulfur was used to make the lake pigments while Sn was detected in tin-based cochineal lake paint out. In comparison, XRF measurements of the samples prepared with just the dyes in the traditional Chinese manner did not detect any S even though alum-glue solution was used to fix the paint layers. This is presumably because the amount of alum present in the 3 thin washes is very low. These results show that the presence of S can be used to distinguish between alum-based lake pigments and dyes applied in the traditional Chinese method. Note that XRF used in open air without helium purge cannot easily detect Al except in the thicker layers of paint with lake pigments. XRF measurements of the paintings showed that neither Sn nor S was ever detected in the areas where scale insect dyes were found, thus ruling out any alum or tin based lake pigments and narrowing it down to just lac or cochineal dyes.

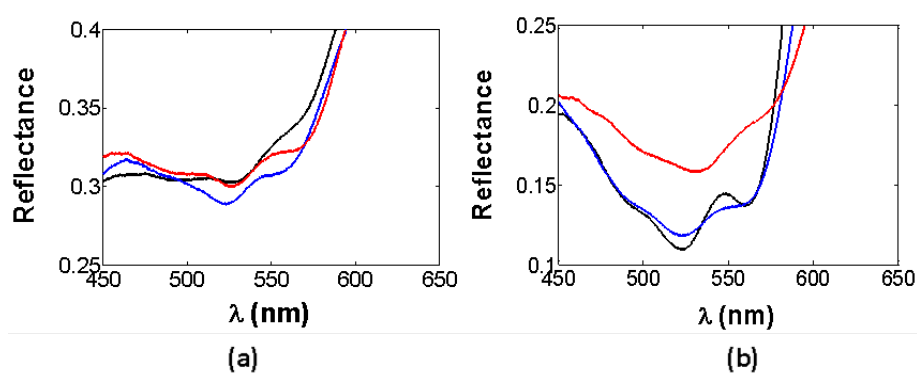

Fig 11. Spectra from the paintings with spectral features of scale insect dyes (black curves) compared with KM fit using lac dye in animal glue painted in the traditional Chinese way (red curves) and using cochineal dye in animal glue (blue curves) combined with vermilion and lead white for a) painting no. 7791-8 (V&A collection) and b) painting no. D82-1886 (V&A collection).

It was observed that in the paintings from the two collections, the reflectance spectra with the two absorption features between 500 and 600 nm were either consistent with lac or cochineal dye in combination with other pigments found in the same region using the KM model (Fig. 11). Figure 11a shows an example where lac dye fitted the measured spectrum better than cochineal dye when KM model fit was employed taking into account of the other pigments that were found in the same area. Figure 11b shows an example where cochineal dye gave better KM fit to the spectrum. Simply comparing the positions of the absorption features with those found in the spectra of the cochineal and lac samples would not give the correct identification as the spectral shape of the other pigments in the measured area can affect the positions of the absorption lines. For example, simulations using the KM model show that the addition of vermilion can shift the central wavelength of the absorption features to shorter wavelength and the addition of indigo can shift the central wavelength to longer wavelength by as much as 4 nm. On the paintings, the scale insect dyes were often found in combination with pigments such as indigo, lead white, vermilion, red ochre, gamboge and malachite. This tentative separation of the scale insect dyes into 2 groups would not have been possible without studying a large sample.

Confirmation of the identification of lac and cochineal can be obtained in the future by surface enhanced Raman spectroscopy [38-39]. The technique is micro-invasive as it would require removing at least a fibre from the painting.

*3.2.3 Palette*

A sample of more than 600 coloured areas was examined. The analysis of pigments on the watercolours has revealed that traditional artists' materials were usually used on the paintings. These include natural minerals, such as azurite and malachite, but also manufactured pigments such as lead white and red lead. The identification of blue, green and red pigments was relatively easy using the multimodal approach. As discussed in the section above, yellow pigments were the most difficult as in most cases yellow organic dyes were used. Table 3 summarises the pigments identified in this study.

Table 3 Palette

| Pigment Colour | Pigments detected |
|---|---|
| *Blue* | *azurite, indigo, Prussian Blue, ultramarine[a], smalt[b]* |
| *Green* | *malachite, botallackite[b]* |
| *Red* | *vermilion, red ochre, red lead, scale insect dyes* |
| *Yellow* | *yellow ochre, gamboge, realgar, chrome yellow[b]* |
| *White* | *lead white* |
| *Gold* | *shell gold* |
| *Silver* | *silver* |

*aDetected on one painting only. Microscope observation shows that it's most probably synthetic given that its particles are very small, rounded and uniformly coloured.*
*bDetected only on one paintings*

## *3.3 Identification of drawing material*

The NIR spectral bands in the PRISMS system were used for a preliminary examination to determine which paintings have preparatory drawings. For more detailed examination and identification of the drawings, the OCT system was used, providing a much higher contrast and spatial resolution. Among the examined paintings there were many with drawings that were uncontaminated by the paint and hence providing the opportunity for Raman measurements. Figure 12 shows a painting with pencil drawings. The high resolution OCT image was averaged over a range of depth slices that showed the drawing material to be of a solid substance. Raman detected graphite which confirmed that it was drawn with a graphite pencil which is in the western tradition. In contrast, Fig. 13 shows an ink drawing that delineates the edge of the table. Both PRISMS and OCT images show that it was drawn with a liquid substance while the Raman measurement found it to be amorphous carbon consistent with the composition of Chinese ink.

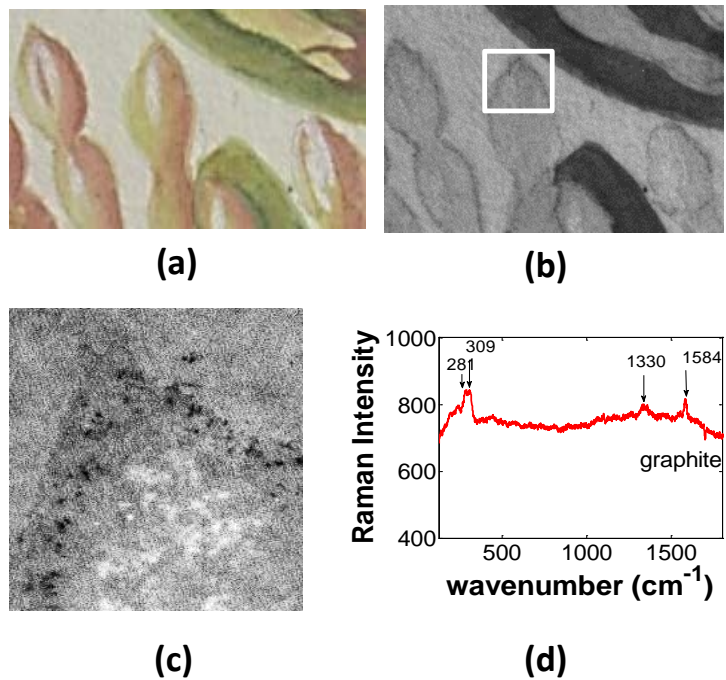

Fig 12. a) Colour image of a part of painting no. 35730 of RHS collection, b) 880nm PRISMS image of the drawings, c) OCT enface image of the boxed area in b) showing the drawings to be of a solid substance and d) the corresponding Raman measurement that identified graphite.

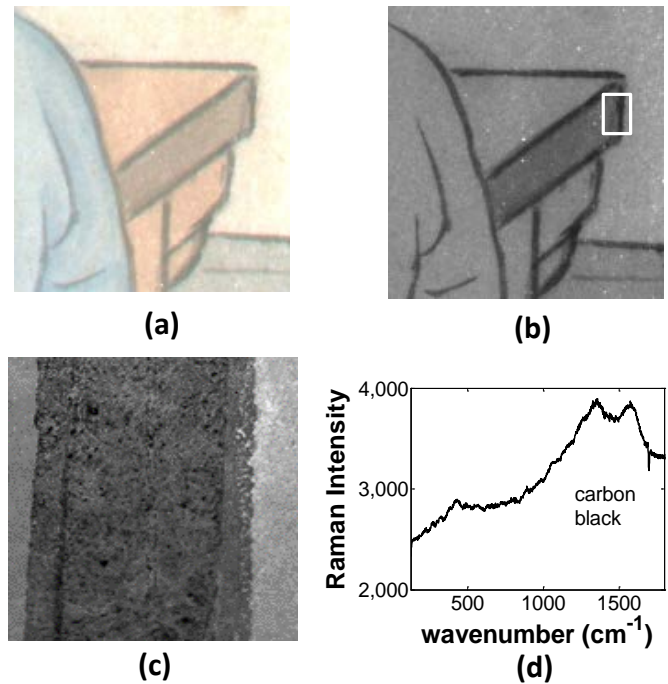

Fig 13. a) Colour image of part of painting no. 7791-12 of V&A collection; b) 880nm PRISMS image of the drawing, c) OCT enface image of the boxed area in b) showing drawing to be of a liquid substance and d) the corresponding Raman spectrum that identifies carbon black.

3.4 *Identification of substrates*

All paintings from the RHS Reeves collection were painted on paper but it is known from the watermarks that some of the papers originated in England. The watermarks indicate 'Whatman' and 'M. J. Lay' with dates consistent with Reeves documentation, that is pre-1830s. In contrast none of the paintings examined in the V&A collection had watermarks. Six of the paintings from the V&A collection were painted on textile, possibly silk. OCT was used to determine which paper substrate was likely to be Western or Chinese. It provides a rapid, non-contact evaluation of the light scattering properties, thickness and fibre length of the paper without having to de-mount the painting.

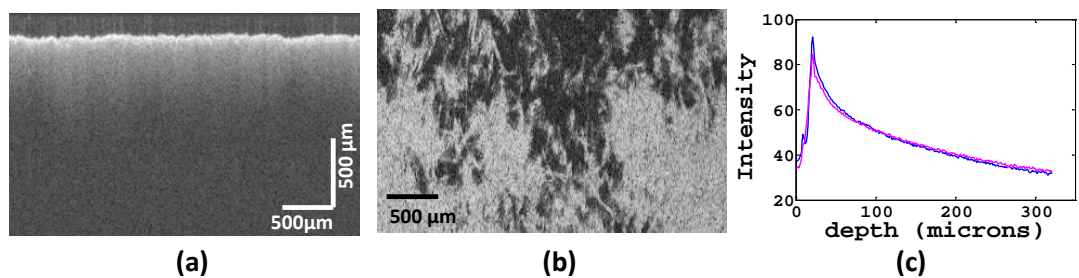

Fig 14. a) OCT cross- section and b) enface image of the English paper (M. J. Lay) substrate of painting no. 36643 (RHS collection) using the OCT system; c) comparison between the depth profile of the backscattered light intensity of the M. J. Lay paper (magenta line) with that of the Whatman paper (blue line).

Two different groups of paper were identified. In the first group the paper fibres were short and highly scattering. An area of unpainted paper was scanned by the OCT to obtain a series of cross-sections (Fig. 14a) in an image cube which was sliced in the direction parallel to the painting surface. Image slices near the surface showed the paper fibres to be short and densely packed (Fig. 14b). An average backscattered light intensity profile as a function of depth (Fig. 14c) was obtained by averaging the depth profiles in one cross-section scan (Fig. 14a). A paper with a watermark clearly identifying it to be a Whatman paper was used as a reference for comparison. Fig. 14c shows that the paper with a watermark identifying it to be a 'M J Lay' paper has a similar light scattering property as the reference Whatman paper. The second group consisted of papers with long fibres (typically at least 5 times longer than those in the first group) consistent with Chinese paper. In many cases, this kind of paper was semi-transparent under the OCT allowing both the paper support and the backing paper to be seen (Fig. 15a). These papers are typically thin with thickness of 100-300 µm assuming an average refractive index of the paper to be given by that of cellulose which is 1.55. It is important to note that OCT measures optical thickness which is the physical thickness multiplied by the refractive index. Fig. 15b shows the long fibre of a semi-transparent thin paper in an OCT image slice near the surface parallel to the surface of the painting. The low light scattering property of the paper was confirmed through the comparison of the slope of the average depth profile of the backscattered light with that of the reference Whatman paper (Fig. 15c). Increased multiple scattering makes the slope shallower. Characteristics such as long fibre, thin paper and low scattering properties are correlated in the same way as short fibre, thick paper and high scattering properties are correlated.

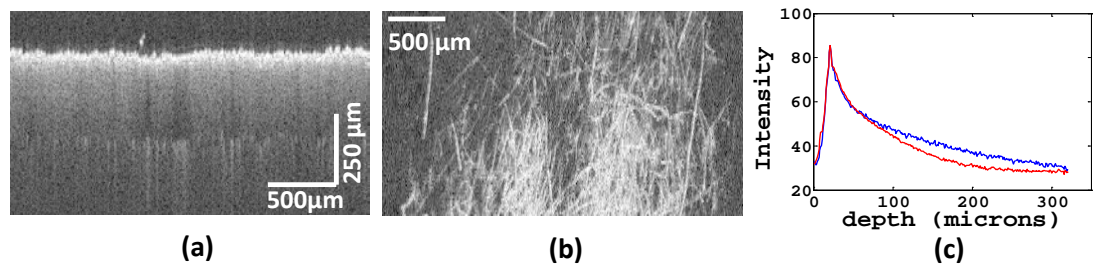

(a)     (b)     (c)

Fig 15. a) OCT cross-section and c) enface image of the long fibre paper substrate of the painting no. 67107 (RHS collection); c) Comparison between the depth profile of the backscattered light intensity of the long fibre paper (red curve) with that of Whatman paper (blue curve).

The two paintings on 'silk' examined by the OCT showed a plain weave (Fig. 16). The threads are not twisted and there is very little spacing between the threads which is why scattering is strong in the OCT image. The spacing between the warp threads and weft threads are even which is different from the conventional weave for silk hanging scrolls [17]. Further analysis is needed to confirm beyond doubt that the textile is indeed silk.

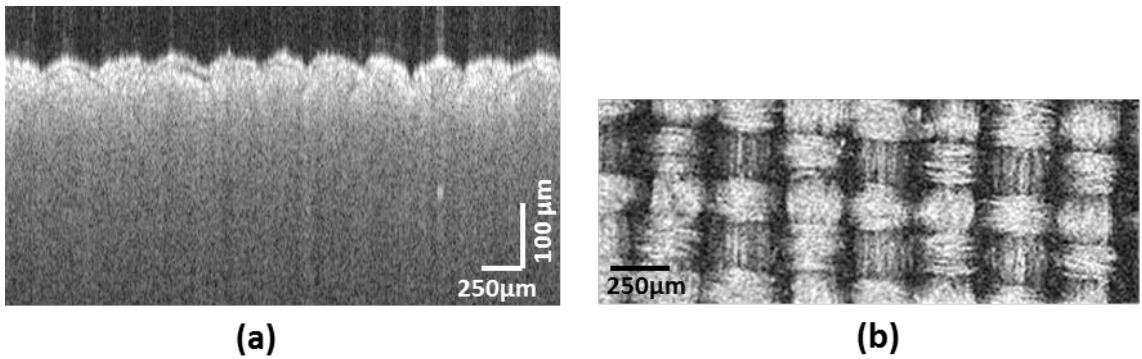

Fig 16  a) OCT cross-section and b) enface image of silk substrate of painting no. 7790-7 (V&A collection)

*3.5 Identification of sizing technique and filler content*

The XRF analysis of blank areas of the substrates showed the presence of sulfur in all the textile substrates. For traditional Chinese silk painting and paintings on paper, it is common to prepare the substrate with a wash of a solution consisting of alum and animal glue [17-18]. Since the XRF was used with an open beam in air without helium purging, it was only sensitive to elements with atomic number Z>14. Unless a huge amount of Al is present, it will not be able to detect the element. However, it is easier to detect the S in alum.

Sulfur was detected in nearly all the paper substrates identified as having long fibre (i.e. likely to be Chinese paper) in the V&A collection, which indicates that the paper substrates were most likely prepared in the traditional manner with a solution of alum and animal glue. It is interesting to note that S was not detected in any of the areas with red dyes (Section 3.2.2) even when they were painted on such sulfur containing papers. The red dyes were always found in combination with pigments such as lead white (Pb) and vermilion (Hg) which can mask the sulfur signal from the paper substrate. In the V&A group of paintings, there were only 2 paintings identified with short fibre papers (i.e. likely to be Western paper) and one of them was examined with XRF but S was not identified indicating that the Western type of paper was treated differently.

In the RHS group of paintings, there were 11 paintings with their substrates examined by XRF and S was not detected in any of the paper substrates except for one, regardless of whether the paper was of the Western or Chinese type. The only paper with S detected was one of the Chinese type where it had long fibres and was semi-transparent.

The papers in the two collections that were identified to be Western were not sized in the traditional Chinese way. It appears that while the Chinese papers were sized in the traditional Chinese way in the V&A group of paintings, most of the RHS paintings were not.

In contrast with the V&A paintings, the RHS paintings had elements other than Ca, K and Fe detected in the substrate. Copper was detected in all the Whatman papers. Both Cu and Ti were detected in the M J Lay paper. In the RHS group of paintings, Cu and sometimes Zn, Pb and Ti were detected in all the other papers regardless of whether they were classified as Chinese or Western. This indicates that there is a fundamental difference in the papers used for the RHS botanical collection compared with those in the V&A collection. Extra elements were detected in the RHS papers except for one. The only exception in the RHS group of paintings was one which had all the characteristics of a Chinese paper (i.e. long fibre and semi-transparent) and was sized the Chinese way. The two paintings in the V&A collection with additional elements detected was one that had Ti and not sized the Chinese way and another that had Cu but sized the Chinese way. The extra elements detected are likely to be due to a difference in the filler material for the papers or in the composition of the size.

None of the Western papers in the two collections were sized using the Chinese method, which is not so surprising. The V&A paintings were almost all on Chinese paper and they appear to be sized in the traditional manner, which is again not surprising. However, the finding that even the papers classified as Chinese in the RHS collection appeared not to be sized in the usual Chinese way was unexpected. In addition, extra elements were detected in the RHS papers regardless of whether they were classified as Chinese or Western.

## **4 Conclusions**

The painting materials of the two collections of Chinese watercolours were analysed non-invasively, using a holistic approach involving VIS-NIR multispectral imaging, high spectral resolution VIS-NIR spectroscopy with FORS, micro-Raman spectroscopy, XRF spectroscopy and OCT imaging. The imaging and spectroscopic techniques complement each other both in terms of their suitability for the identification of different types of materials and in terms of spatial resolution and field of view.

VIS-NIR reflectance spectroscopy was shown to be capable of detecting single and multiple pigments either in a mixture or in layers by using an algorithm based on the Kubelka-Munk model. It was effective at identifying most of the blue, green and red pigments as they have distinctive spectral characteristics in the VIS-NIR. However, it was ineffective at distinguishing between the majority of the yellow pigments. FORS is the most effective non-invasive technique at identifying scale insect dyes. In the context of these paintings, a tentative distinction between lac and cochineal dyes seemed possible with the large dataset collected, although further confirmation possibly through surface enhanced Raman spectroscopy is required. The XRF provides elemental identification for elements with Z>14 and is therefore mostly ineffective for organic materials. Micro-Raman has the advantage of giving highly specific molecular identification on the scale of individual particles and therefore reducing interference from surrounding materials. However, the trade-off is the small microscopic spot size which does not always give a representative view of the paint composition even with the help of visual examination through the microscope. The Raman signal is often masked by fluorescence from organic materials, though the technique can detect some organic pigments such as gamboge and indigo. VIS-NIR spectroscopy either by multispectral imaging or FORS gives a spatially more representative view of the painted area compared with XRF and Raman due to the larger spot size or field of view. Therefore it was necessary to use all 3 techniques for pigment identification. Binder identification in the SWIR was hindered by the absorption features of the paper substrate. In future, a portable FTIR might be used to assist in binder identification, although it may still be challenging given the small amount of binder present.

Multispectral imaging (NIR bands) allowed rapid visualisation of the preparatory drawings, OCT imaging gave detailed high resolution and depth-resolved, high-contrast images of the drawings in smaller areas to distinguish between drawings made with a solid or a liquid substance, while micro-Raman spectroscopy confirmed the drawing material to be either graphite (pencil) or carbon (ink).

OCT imaging was able to distinguish between papers with long fibres that were thin and semi-transparent (likely to be Chinese) from those with short fibres that were thick and highly scattering (likely to be Western). It also provided rapid quantitative measures of the fibre length and the thickness of the Chinese papers non-invasively, without the need to de-mount the paintings. XRF revealed whether the papers were sized using the traditional Chinese method and whether the size and filler materials for the papers were different.

The palette of the RHS paintings and the V&A paintings are largely similar where mostly traditional Chinese pigments (natural or synthetic) were used. While it is known that cochineal and Prussian blue were imported in significant quantities to China during the period when the RHS watercolours were painted, there is so far no evidence for Prussian blue on any of the paintings analysed. Prussian blue was however detected in 8 of the V&A paintings. Scale insect dyes appear to have been used on nearly every painting from the RHS and the V&A. It seems that both lac and cochineal were used, though cochineal seems to have been used more frequently.

In terms of the drawing material, types of paper and painting techniques, the RHS collection of Chinese botanical watercolours is very different from the paintings in the V&A collection. This is not too surprising since the RHS paintings were commissioned by the Society to serve as a plant catalogue, which is very different from the purpose of the export paintings in the V&A collection. Graphite pencil was used for preparatory sketches on the majority of the RHS paintings while Chinese ink was used on the V&A paintings. The majority of the paintings in the RHS collection were on short fibre papers while nearly all of the V&A paintings were on long fibre papers. All but one of the V&A paintings were sized with alum while only one of the RHS paintings was sized with alum. Elements such as Cu, Zn, Ti or Pb were detected on nearly all of the RHS papers regardless of whether they were long or short fibre papers. These elements were largely absent from the V&A papers.

A micro-destructive technique, microfade spectrometry, was used to assess the vulnerability of the paintings to light exposure. With the exception of a realgar containing paint, all paints and paper substrates were found to be more stable than ISO Blue Wool 3 with the paints being in general more stable than the paper substrates. These results will inform decisions on exhibition and storage conditions for these paintings.

A follow-up paper will address in detail the significance of the scientific results in light of the contemporary historical records.

**Acknowledgements**

This project was funded by the UK AHRC/EPSRC Science and Heritage Programme, research development award AH/K006339/1. We are grateful to Jo Kirby of National Gallery (London) for reference samples of various lake pigments and useful discussions, Hongxing Zhang of the Victoria and Albert Museum for initiating the project. We thank students from Nottingham Trent University, Chris Dilley, Romain Le Charles and Elias Cheyroux for their assistance.

**References**

[1] C. Miliani, F. Rosi, B. Brunetti, A. Sgamellotti, Accounts of Chemical Research **43**(6), 728 (2010)

[2] C. Miliani, F. Rosi, A. Burnstock, B.G. Brunetti, A. Sgamellotti, Appl. Phys. A **89**, 849 (2007)

[3] P. Westlake, P. Siozos, A. Philippidis, C. Apostolaki, B. Derham, A. Terlixi, V. Perdikatsis, R. Jones, D. Anglos, Anal. Bioanal. Chem. **402**, 1413 (2012)

[4] S. Bruni, S. Caglio, V. Guglielmi, G. Poldi, Appl. Phys. A **92**, 103 (2008)

[5] M. Aceto, A. Agostino, G. Fenoglio, M. Gulmini, V. Bianco, E. Pellizzi, Spectrochim Acta Part A **91**, 352 (2012)

[6] S. Neate, D. Howell, in The Technological Study of Books and Manuscripts as Artefacts: Research Questions and Analytical Solutions, ed. By S. Neate, D. Howell, R. Ovenden, A. M. Pollard, British Archaeological Reports International Series, 2209, (Archaeopress, Oxford, 2011), p. 9

[7] C. Clunas, Chinese Export Watercolours (Victoria and Albert Museum, London, 1984)

[8] C. Bailey, Studies in Conservation **57**(2), 116 (2012)

[9] H. Liang , L. Burgio , C. Bailey, A. Lucian, C. Dilley, S. Bellesia, C. Brooks, C. S. Cheung, Studies in Conservation **59**, S96 (2014)

[10] P. Whitemore, X. Pan, C. Baillie, Journal of the American Institute of Conservation **38**, 395 (1999)

[11] H. Liang, R. Lange, A. Lucian, P. Hyndes, J. Townsend, S. Hackney, in ICOM Committee for Conservation, 16th Triennial Conference Lisbon, (Lisbon: Critério - Artes Gráficas, Lda. 2011), p.1612_882

[12] H. Liang, Appl. Phys. A **106** (2), 309 (2012)

[13] H. Liang , A. Lucian, R. Lange, C. S. Cheung, B. Su, ISPRS Journal of Photogrammetry and Remote Sensing **95**, 13 (2014)

[14] P. Ricciardi, J. Delaney, M. Facini, J. Zeibel, M. Picollo, S. Lomax, M. Loew, Angew. Chem. Int. Ed. **51**, 5607 (2012)

[15] K. A. Dooley, S. Lomax, J. G. Zeibel, C. Miliani, P. Ricciardi, A. Hoenigswald, M. Loew, J. K. Delaney, Analyst **138**, 4838 (2013)

[16] M. Aceto, A. Agostino, G. Fenoglio, A. Idone, M. Gulmini, M. Picollo, P. Ricciardi, J. Delaney, Analytical Methods **6**, 1488 (2014)

[17] J. Winter, East Asian Paintings – Materials, Structures and Deterioration Mechanisms (Archetype Publications, London, 2008)

[18] F. Yu, Chinese Painting Colors: Studies of their Preparation and Application in Traditional and Modern Times, J. Silbergeld and A. McNair, trans. (Hong Kong University Press and University of Washington Press, Hong Kong and London, 1988)

[19] R. Clark, Chem. Soc. Rev. **24**, 187 (1995)

[20] P. Vandenabeele, H. G. M. Edwards, J. Jehlicka, Chem. Soc. Rev. **43**, 2628 (2014)

[21] L. Burgio, in The Technological Study of Books and Manuscripts as Artefacts: Research Questions and Analytical Solutions, ed. By S. Neate, D. Howell, R. Ovenden, A. M. Pollard, British Archaeological Reports International Series, 2209, (Archaeopress, Oxford, 2011), p. 67

[22] H. Liang, M. Gomez Cid, R. G. Cucu, G. M. Dobre, A. Gh. Podoleanu, J. Pedro, D. Saunders; Opt. Express **13**, 6133 (2005)

[23] C. S. Cheung, J. M. O. Daniel, M. Tokurakawa, W. A. Clarkson, H. Liang, Optics Express **23**, 1992 (2015)

[24] E. Alarousu, L. Krehut, T. Prykari, R. Myllyla, Meas. Sci. Technol. **16**, 1131 (2005)

[25] H. Liang, R. Lange, B. Peric, M. Spring, Appl. Phys. B **111**, 589 (2013)

[26] D. Saunders, J. Kirby, National Gallery Technical Bulletin **15**, 79 (1994)

[27] C. Clementi, D. Doherty, P. L. Gentili, C. Miliani, A. Romani, B. G. Brunetti, A. Sgamellotti, Appl. Phys. A **92**, 25 (2008)

[28] I. M. Bell, R. J. H. Clark and P. J. Gibbs, , Spectrochim Acta A **53A**, 2159 (1997)

[29] L. Burgio, R. J. H. Clark, Spectrochim Acta A **57**, 1491 (2001)

[30] M. Aru, L. Burgio, M. S. Rumsey, J. Raman Spectrosc. **45**, 1006 (2014)

[31] B. H. Berrie, Proc. Natl. Acad. Sci. USA **106**, 14757 (2009)

[32] J. Kirby, National Gallery Technical Bulletin **14**, 35 (1977)

[33] M. Leona, J. Winter, Studies in Conservation **46**, 153 (2001)

[34] J. Winter, J. Giaccai, M. Leona, in Scientific Research in the Field of Asian Art: Proceedings of the First Forbes Symposium at the Freer Gallery of Art, ed. By P. Jett, J. Douglas, B. McCarthy, J. Winter (Archetype Publications, London, 2003), p. 157

[35] C. Bisulca, M. Picollo, M. Bacci, D. Kunselman, in the 9[th] International Conference on NDT of Art, www.ndt.net/search/docs.php3?MainSource=65, Jerusalem Isreal, May 2008

[36] G. Favaro, C. Miliani, A. Romani, M. Vagnini, J. Chem. Soc., Perkin Trans. **2**, 192 (2002)

[37] M. V. Canamares, M. Leona, J. Raman Spectrosc. **38**, 1259 (2007)

[38] Y. Strumfels, B. H. Berrie, Studies in Conservation **47**, Supplement-2, 18 (2002)

[39] A. W. Whitney, R. P. Van Duyne, F. Casadio, J. Ramn Spectrosc. **37**, 993 (2006)